\documentclass[lettersize,journal]{IEEEtran}

\ifCLASSOPTIONcompsoc
    \usepackage[caption=false, font=normalsize, labelfont=sf, textfont=sf]{subfig}
\else
\usepackage[caption=false, font=footnotesize]{subfig}
\fi
\usepackage{amsmath,amssymb,amsfonts, amsthm}
\usepackage{mathtools}

\usepackage{textcomp}
\usepackage{stfloats}
\usepackage{url}
\usepackage{verbatim}
\usepackage{graphicx}
\usepackage{cite}
\usepackage{booktabs}
\usepackage{multirow}
\usepackage[flushleft]{threeparttable}
\usepackage[super]{nth}
\usepackage{xcolor}
\usepackage{colortbl}
\usepackage{algorithm}
\usepackage{algorithmic}
\usepackage[caption=false, font=footnotesize]{subfig}
\usepackage{tikz}
\usetikzlibrary{arrows.meta, positioning}
\usepackage{soul}
\setstcolor{red}
\usepackage{rotating}
\newcommand{\blue}[1]{\textcolor{black}{#1}} 

\usepackage{bm}
\usepackage{algorithmic}
\usepackage{algorithm}
\usepackage{array}

\newcounter{thm}
\newtheorem{prop}{Proposition}
\newtheorem{definition}{Definition}
\newtheorem{prob}[thm]{Problem}

\newtheorem{theorem}{Theorem}[section]

\usepackage{mathtools}

\usepackage[acronym]{glossaries}
\makeglossaries
\newacronym{RSR}{RSR}{Request Service Rate}

\newcommand{\rev}[1]{{\leavevmode\color{black}#1}}
\newcommand{\reva}[1]{{\leavevmode\color{black}#1}}
\newcommand{\revb}[1]{{\leavevmode\color{black}#1}}
\newcommand{\revf}[1]{{\leavevmode\color{black}#1}}

\tikzset{every picture/.style={/utils/exec={\sffamily}}}

\begin{document}
\title{Sink Proximity: A Novel Approach for Online Vehicle Dispatch in Ride-hailing}
\author{Ruiting Wang, Jiaman Wu, Fabio Paparella, Scott J. Moura, Marta C. Gonzalez
\thanks{The authors are with the Department of Civil and Environmental Engineering, University of California, Berkeley, CA, 94720 USA. \tt \{rtwang, fabio\textunderscore paparella, smoura, martag\}@berkeley.edu}

\thanks{J.Wu is with the Department of Data Science, City University of Hong Kong, Hong Kong SAR, China. \tt\{jm.wu\}@cityu.edu.hk}

\thanks{
F. Paparella is also with the Department
of Mechanical Engineering, \mbox{Eindhoven University of Technology,} Eindhoven, 5600 MB, the Netherlands. \tt f.paparella@tue.nl.}
}

\markboth{IEEE Transactions on XXXX}%
{Shell \MakeLowercase{\textit{et al.}}: A Sample Article Using IEEEtran.cls for IEEE Journals}

\IEEEpubid{0000--0000/00\$00.00~\copyright~2021 IEEE}

\maketitle

\begin{abstract}

Ride-hailing platforms have a profound impact on urban transportation systems, \blue{and their performance largely depends on how intelligently they dispatch vehicles in real time. In this work, we develop a new approach to online vehicle dispatch that strengthens a platform’s ability to serve more requests under demand uncertainty. }
\blue{W}e introduce a novel \blue{measure} called \emph{sink proximity}, \blue{a network-science-inspired measure that captures how demand and vehicle flows are likely to evolve across the city.}
\blue{By integrating this measure into a shareability-network framework, we design an online dispatch algorithm that naturally considers future network states, without depending on fragile spatiotemporal forecasts.}
Numerical studies demonstrate that our proposed solution significantly improves the request service rate under peak hours within a receding horizon framework with limited future information available.

\end{abstract}

\section{Introduction}
Shared mobility services have greatly transformed urban transportation and provided a convenient alternative to reduce the use of private vehicles \cite{banerjee2019}. The global market size of ride-hailing was 106.66 billion in 2023 and is projected to grow from \$123.08 billion in 2024 to \$480.09 billion by 2032 \cite{market_2023}. 
It is fundamental for these shared mobility services to become both efficient and profitable. 
\blue{P}latform-based operators without vehicle ownership primarily rely on increasing the volume of successfully matched driver–rider pairs to boost revenues \cite{UberProf}.
\blue{A core challenge is the platform's ability to effectively assign} drivers to incoming ride requests, a problem commonly referred to as the \emph{Vehicle Dispatch Problem} (VDP).

From an economic standpoint, the dispatch algorithm significantly influences how effectively the platform balances demand from riders and supply of drivers, thereby reducing driver idle time and lowering passenger waiting time, and optimizing revenue for the platform \cite{qin2021optimizing, castillo2024matching}. A good matching strategy \revb{also influences energy consumption and environmental impact, particularly in electrified fleets where smart coordination with power grids is essential \cite{Marzbani2024,RossiIglesiasEtAl2018b}.}

\revb{Besides financial incentives, there are important environmental and social motivations for improving the VDP. Strategically matching drivers to riders can minimize unnecessary vehicle movements, reduce overall energy consumption, and decrease emissions \cite{Wollenstein-BetechSalazarEtAl2021,SalazarLanzettiEtAl2019,jin2024, li2024a}. Such strategies also help mitigate traffic congestion during peak hours \cite{wei2022, ke2020}. Thus, enhancements to the VDP can support wider sustainability goals by making transportation systems cleaner, more efficient, and more resilient.}

This paper proposes a novel network science approach to the VDP, which is formulated as a network flow problem and solved in a receding horizon control (RHC) fashion. By incorporating a specialized \blue{measure} derived from network science, we demonstrate how the request service rate (i.e., the fraction of riders served) can be improved compared to a standard RHC formulation.

\IEEEpubidadjcol

\subsection{\rev{Vehicle Dispatch in Ride-hailing}}

With the increasing size of the market and the large number of drivers and riders in the platform, \revb{assigning drivers to real-time ride requests becomes increasingly complex.} Besides the \revb{sheer} scale of the problem, the inherent unpredictability of future demand \revb{adds another layer of difficulty. Research on VDP typically splits into two main methodological approaches:}

\subsubsection{\revb{Optimization-Based Methods}}

Combinatorial optimization algorithms often present the problem as \revb{a variant of the vehicle routing problem.} This format is more flexible and allows for customized constraints. 
To address the lack of information in online optimization and uncertainty on the demand side, studies have used several approaches, such as Bayesian framework \cite{zhang2017}, multi-stage stochastic optimization \cite{lowalekar2018online}, sampling \cite{xu2021network}, etc.
The primary challenge in employing combinatorial optimization techniques also lies in the computational complexity, which is crucial for online implementation in large-scale systems.

\subsubsection{\revb{Machine Learning and Reinforcement Learning}}

In recent years, studies have based their work on \blue{machine learning and }reinforcement learning (RL) models, to handle uncertain and fast-evolving environments, where dispatch decisions must be updated within seconds \cite{RL_wild, Deep_RL, Combi_RL}.
These methods can be adept at learning from data in real time but often struggle with transferability if new conditions differ significantly from training data \cite{tang2021a, liang2022}. \blue{Readers can refer to these survey papers \cite{wen2024a, liu2022} for a comprehensive overview of how different methods are utilized in different problems in ride-hailing, including but not limited to VDP.}

A middle-ground solution \revb{uses network flow formulations, which can be solved efficiently by exploiting graph-theoretic
properties}~\cite{ford1956, orlin1996, tarjan1997}. 
A variety of complex and realistic situations can be implemented by modifying the structure, and \revb{recent studies also incorporate ideas from learning frameworks into these classical algorithms to balance computational tractability with adaptive capabilities}~\cite{zhan2016, vazifeh2018addressing, xu2021network, wu2021mobility}.

\blue{Regardless of the algorithms used}, the choice of objective can also vary, reflecting the diverse needs and constraints of ride-hailing platforms. \blue{To highlight a few directions, besides matching algorithms, }some studies focus on: dynamic pricing~\cite{yan2020}, electric fleets \cite{RossiIglesiasEtAl2018b}, interaction with public transit \cite{Wollenstein-BetechSalazarEtAl2021}, 
the fairness among drivers \cite{shi2021}, and evaluation of environmental, social, and economic benefits of ride-hailing \cite{jin2024}. 

\subsection{Network Science for Vehicle Dispatch}

\revb{Network science offers a variety of \blue{measures} (e.g., in-degree, betweenness, closeness \cite{freeman1977, bavelas1950, katz1953}) to analyze the structure of complex networks. However, these \blue{measures} typically assume static graphs. Real-world shareability networks in ride-hailing are dynamic, meaning nodes (ride requests) and edges (potential matches) evolve over time. Although some work has studied \blue{measures} for dynamic networks (e.g., communication \cite{buechel2013}, social networks \cite{elmezain2021}), no existing research has linked network science \blue{measures} to dynamic shareability networks in ride-hailing.}

\revb{Addressing this gap, specifically, how future network states may influence current matching decisions—can yield valuable insights and more efficient online algorithms for the VDP.}
This work aims to fill the gap by designing a tool that can be leveraged in network flow problems in a dynamic \revb{network} setting. In particular, we focus on the temporal evolution of the shareability network and design a novel network science \blue{measure}, \emph{sink proximity} (SP), that captures the potential importance of each node in the network flow. 

\subsection{Statement of Contribution}
\revb{This study} utilizes the network topology of a shareability network in ride-hailing problems through the application of network science to enhance the efficiency of ride-hailing platforms. 
We claim the following contributions:

\subsubsection{\revb{Novel \blue{Measure}—Sink Proximity}}

We introduce \emph{sink proximity} in the context of network flow problems, \revb{which} quantifies the longest path distance from any given node to the sink node in a network flow. Within shareability networks, this \blue{measure} \revb{helps to account for downstream opportunities and} incorporates elements of future information regarding the network's expansion.

\subsubsection{\revb{Efficient Computation in a DAG}}

We demonstrate \revb{how} \emph{sink proximity} can be efficiently approached as a single source longest path problem in a directed acyclic graph (DAG), which can be solved in polynomial time. \revb{This ensures that real-time ride-hailing systems can feasibly integrate the \blue{measure} into online algorithms.}

\subsubsection{\revb{Application to Vehicle Dispatch}}

\revb{We incorporate sink proximity into a network-flow-based receding horizon framework \blue{(RHC-SP)}. Simulation results show how this approach boosts the request service rate relative to standard RHC solutions.} This algorithm represents a significant advancement in optimizing the dispatch process through the application of network science principles.

\revb{Without} loss of generality, we focus on the standard VDP\revb{, though the proposed method can be adapted to variations, such as profit maximization \cite{PaparellaHofmanEtAl2024} or energy-aware dispatching \cite{RossiIglesiasEtAl2018b}. }The objective of this study is to enhance the operational efficiency of ride-hailing platforms by optimizing the \gls{RSR}, i.e., the number of orders served normalized by the overall number of orders of the vehicle dispatch problem.

\section{Problem Formulation}
In this section, first we formally define the orders and drivers. Then we define the dispatching problem as an integer linear problem \blue{as Problem~\ref{prob:one}}. Then, we leverage a vehicle-shareability network approach to transform the \blue{Problem~\ref{prob:one}} into a directed acyclic graph formulation (\blue{Problem~\ref{prob:two}}). This allows us to use \blue{polynomial-time machinery developed} for minimum/maximum-cost flow problems \blue{to solve the dispatching problem efficiently}. 

\subsection{\rev{Vehicle} Dispatch Problem Formulation}
The dispatch problem within a ride-hailing platform involves three key participants: the platform itself, drivers, and passengers. The platform aggregates a collection of {orders and drivers, which we define in the following.} We refer to this challenge as the ``vehicle dispatch problem''. 

\begin{definition}
We define an order as a tuple $\mathcal{O}_i = (t_{i}^{p},t_{i}^{d},o_{i},d_{i}) \in \mathcal{T} \times \mathcal{T} \times \rev{\mathcal{Y}} \times \rev{\mathcal{Y}}$, where $\mathcal{T}$ is the simulation time period and $\rev{\mathcal{Y}}$ is the set of \rev{locations} in which an order can start and end the trip. Then, $t_{i}^{p}$ and $t_{i}^{d}$ are the anticipated pick-up and drop-off times, while $o_{i}$ and $d_{i}$ indicate the origin and destination locations, respectively. Last, we define the set of orders $\mathcal{O}:= \{\mathcal{O}_i\}_{i\in \mathcal{N}}$, where $\mathcal{N}=[1,2,...,N]$ is the set of indices of such orders.
\end{definition}
It is assumed that if a passenger $i$\footnote{\rev{For the reminder of the paper, we use interchangeably $\mathcal{O}_i$ and $i$ to indicate the $i$-th order, and $\mathcal{D}_k$ and $k$ to indicate the $k$-th driver.}} places an order at time $t_{i}^{s}$ with a maximum waiting time of $t_{i}^{w}$, then the expected pick-up time is $t_{i}^{p} = t_{i}^{s} {\leq} t_{i}^{w}$. 

\begin{definition}
    We define a driver as a tuple $D_k = (t^r_k,p^t_k)\in \mathcal{T} \times \rev{\mathcal{Y}}$, where $t^r_k$ indicates the time they begin to accept orders, and $p^t_k$ signifies their location at time $t^r_k$. Last, we define a set of drivers $\mathcal{D}:=\{\mathcal{D}_k\}_{k \in \mathcal{M}}$, where $\mathcal{M}=[1,2,...M]$ is the set of indices of such drivers.
\end{definition}

Importantly, drivers have a maximum idling time, formally defined by $t^{idle}_{k} \leq T_{p}, \forall k \in \mathcal{M}$.

The platform seeks to maximize the number of served orders while satisfying the maximum idle time constraints. {The resulting} integer linear programming (ILP)\cite{wu2021mobility} is {defined in the following}.

\begin{prob}(Vehicle Dispatch Problem)\label{prob:one}

Given a set of orders $\mathcal{O}$ and a set of drivers $\mathcal{D}$, the decision variables $x_{ij}^k,s^k_i,t^k_i$ that maximize the \revf{\gls{RSR},} result from
\begin{align}
    \max \quad& \frac{\sum\nolimits_{i, j, k} x^k_{ij} -1}{N}\quad&\\
    \label{c1}
    \mbox{s.t.}\quad& \sum\nolimits_{k,j}x^k_{ij} \leq 1, \quad & \forall i,\\
    \label{c2}
    &\sum\nolimits_{j}x^k_{ij} - \sum\nolimits_{j}x^k_{ji}= s^k_i-t^k_i, \quad &\forall i, k,\\
    \label{c3}
    &\sum\nolimits_{i} s^k_i \leq 1, \quad\sum\nolimits_{i} t^k_i \leq 1, \quad & \forall k,\\
    \label{c5}
    &x_{ij}^k \leq \hat{x}^k_{ij}(T_p),\quad x_{ij}^k, s^k_i, t^k_i \in \{0,1\} \quad & \forall i, j, k,
\end{align}
\end{prob}
\revf{where the objective is the \gls{RSR} (the $-1$ takes into account the trips to the virtual nodes)}, and where $\hat{x}_{ij}^k(T_p)$ is an upper-bound of variable $x_{ij}^k$, and it is defined as follows:
{
\begin{align}
  \!\hat{x}^k_{ij}=\!\left\{
\begin{aligned}
&1,\;\min(v(t_i^d-t^p_j),vT_p)\!\geq\!L(d_i,o_j)\; \land \; t_i^p\!\geq\!t^r_k, \\
&0,\;{\mathrm{otherwise}}.
\end{aligned}
\right.  
\label{Ifijk}
\end{align}
}

The decision variables of {Problem~\ref{prob:one} are 
$x_{ij}^k\in\{0,1\}$, $s_i^k\in\{0,1\}$, and $t_i^k\in\{0,1\}$}. They respectively denote i) whether driver $k$ picks up order $j$ after serving $i$; ii) whether order $i$ is the first or iii) the last order for driver $k$ in the operational period. 
{The upper-bound $\hat{x}_{ij}^k (T_p)$} represents the {compatibility} between order $i$ and $j$, and driver $k$ with maximum idle time $T_p$, and indicates {if} order $j$ can be reached after a driver finishes serving order $i$. {The parameter} $v$ denotes the average speed of the vehicles, while $L(d_i, o_j)$ denotes the distance between location $d_i$ and $o_j$. The constraints (\ref{c1}) indicate that one order can only be fulfilled by one driver. Meanwhile, the constraints (\ref{c2}) allow each driver to serve a sequence of orders, except for their first or last order. Additionally, the constraints (\ref{c3}) ensure that each driver has at most one first order and last order.  Finally, the constraints (\ref{Ifijk}) eliminate unfeasible connections in Eq. (\ref{c5}).

\rev{This stylized vehicle dispatch problem can be solved with a set of heuristics that provide good feasible solutions \cite{bertsimas2019online}.}

\revf{Note that the platform can chose to not assign a vehicle to a particular order if it is not convenient. In that case, the order remains unfulfilled, and the user can either leave the platform or ask again for the service. However, there is no guarantee that an user will be served.}

\subsection{Network Flow Formulation} \label{sec:network_formulation}

\begin{figure*}[hbtp]
\vspace{-6pt}
\centering
    \subfloat[Geographical representation of a set of orders $\mathcal{O}$.]
    {\includegraphics[height=.29\linewidth,trim={0 0 0 0}]{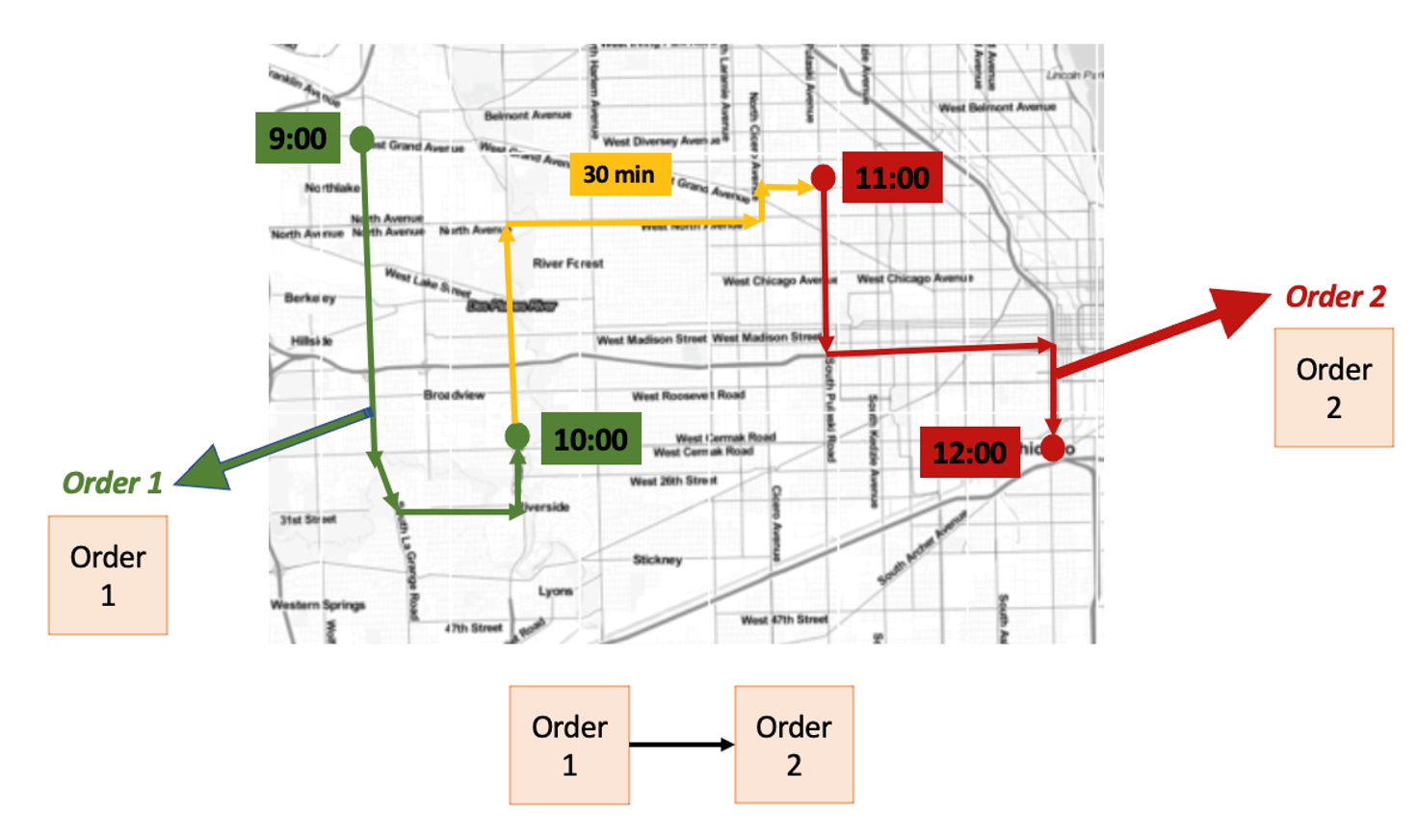}}   \quad \quad
    \subfloat[Simplified shareability network of the set $\mathcal{O}$ with sink.]
    {\includegraphics[height=.29\linewidth,trim={0 9 0 0}]{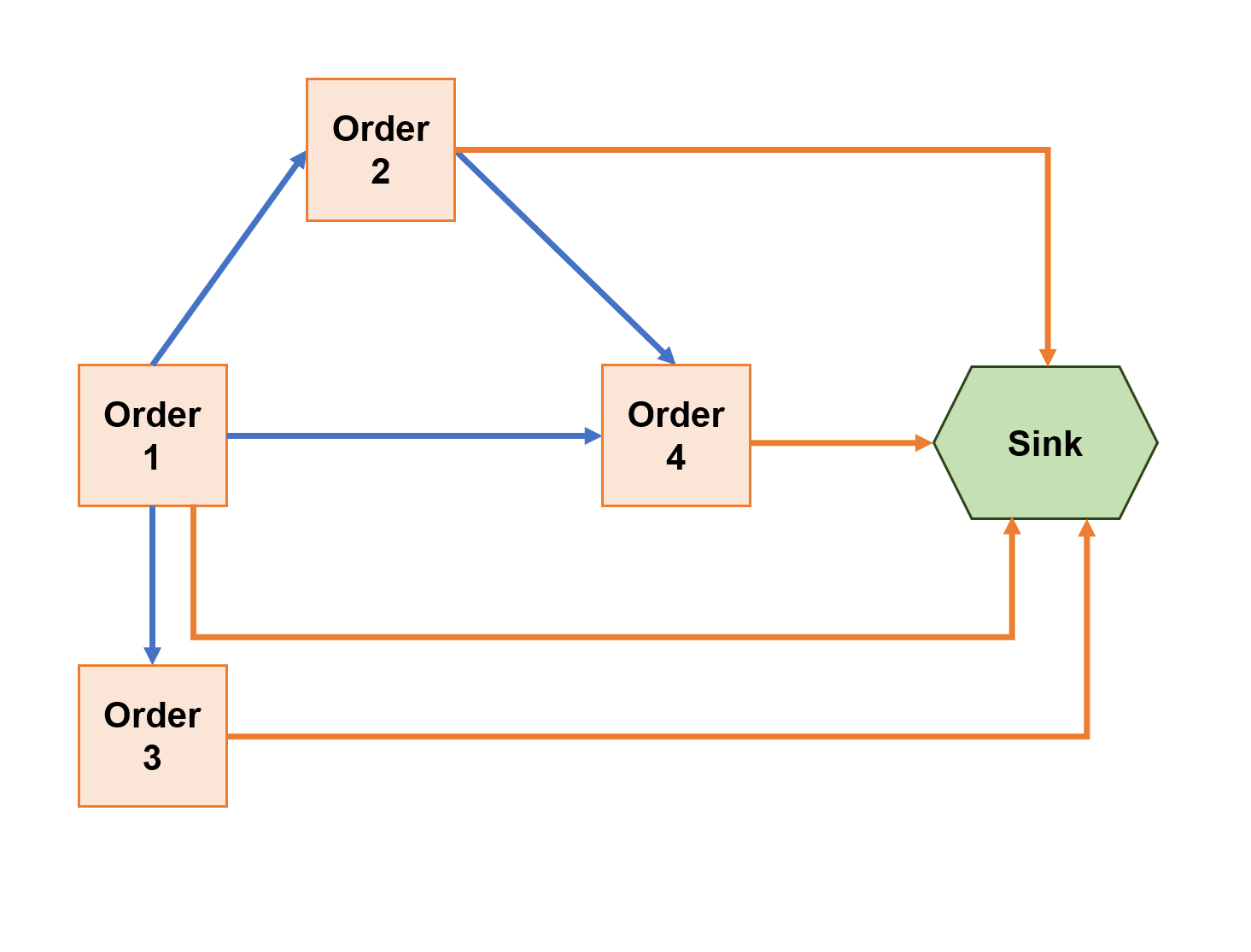}} \\
    \subfloat[Order node structure with internal link\protect\footnotemark.]
    {\includegraphics[height=.2\linewidth]{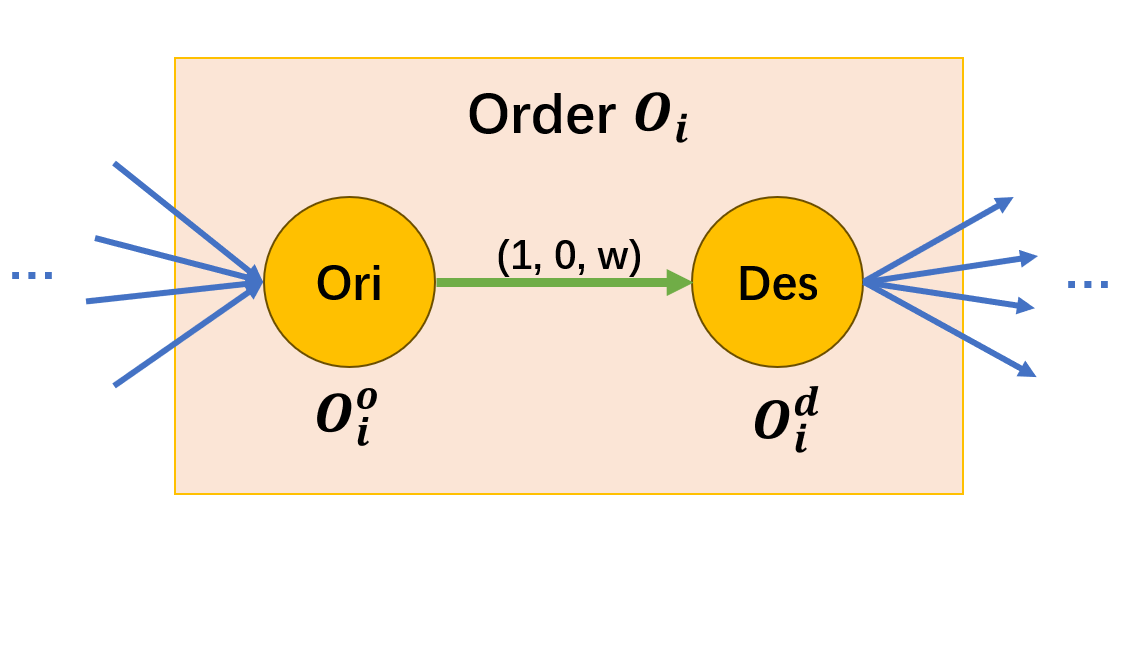}}  \quad 
    \subfloat[Extended shareability network of a set of orders $\mathcal{O}$, drivers $\mathcal{D}$, source and sink.]
    {\includegraphics[height=.3\linewidth,trim={0 0 0 0}]{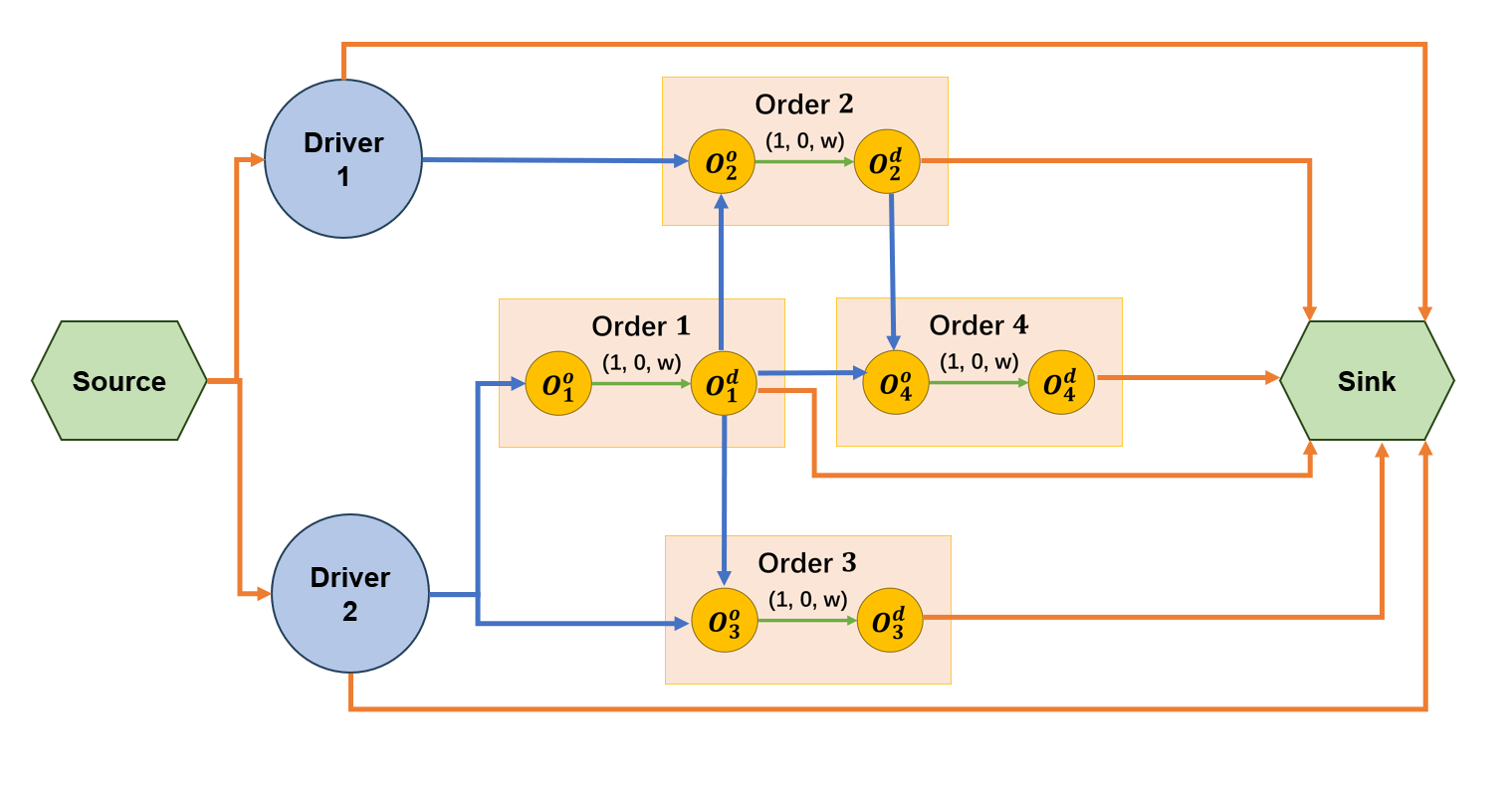}}  \\
    \includegraphics[width=.7\linewidth,trim={0 0 0 0}]{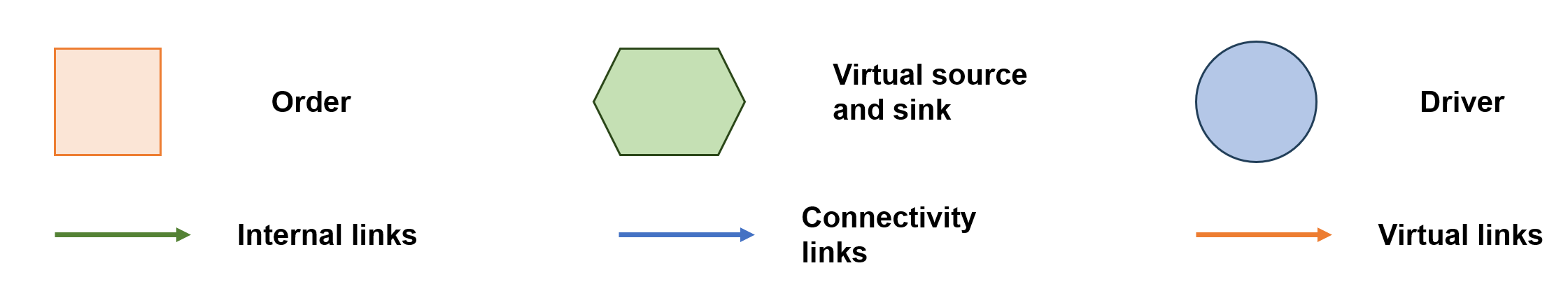}
\caption{Geographical representation, and shareability networks for a set of orders $\mathcal{O}$ and drivers $\mathcal{D}$. The type of links is detailed at the bottom of the figure. }\label{fig:toy}
\vspace{-6pt}
\end{figure*}

To cast Problem~\ref{prob:one} as a network flow problem, we define an extended shareability network, $\mathcal{G}^\mathrm{ext} = (\mathcal{V}^\mathrm{ext},\mathcal{E}^\mathrm{ext})$. \rev{The} set of nodes is defined as $\mathcal{V}^\mathrm{ext} := \rev{\mathcal{O}} \cup \mathcal{D} \cup s \cup t$, and it is the union \rev{of the order nodes $\mathcal{O}$, driver nodes $\mathcal{D}$, virtual source $s$ and virtual sink $t$}.
Each order node in the extended shareability network $\mathcal{G}^\mathrm{ext}$ is further detailed by two nodes, \rev{which represent the origin and destination nodes}, as shown in Fig.~\ref{fig:toy}(c) and in Fig.~\ref{fig:toy}(d).
\rev{The set of links is defined as $\mathcal{E}^\mathrm{ext} := \mathcal{E}^\mathrm{v} \cup \mathcal{E}^\mathrm{c} \cup \mathcal{E}^\mathrm{i} $.} 

Thus, there are three types of links in the extended shareability network: i) the virtual links that connect the virtual node $s$ to drivers, or orders/drivers to $t$; ii) the connectivity links between drivers and orders, and between orders; and iii) the internal links inside each order node.

\revb{The problem can be represented as the following maximum-cost flow problem (MCFP).} 

\begin{prob}(\revb{Maximum Cost Flow Representation of VDP})\label{prob:two}

\revb{Given a directed flow network $\mathcal{G}^\mathrm{ext} = (\mathcal{V}^\mathrm{ext},\mathcal{E}^\mathrm{ext})$. For each edge $(i,j) \in \mathcal{E}^\mathrm{ext}$, a flow $f_{ij}$,  an upper bound $u_{ij} = 1$, a lower bound $l_{ij} = 0$, and a weight $w_{ij}$ are defined.}
\revb{
\begin{align}
    \max_{f_{ij}} \quad& \sum_{(i,j) \in \mathcal{E}^\mathrm{ext}}{w_{ij}f_{ij}} \quad\\
    \mbox{s.t.} \quad & l_{ij} \leq f_{ij} \leq u_{ij}, \quad & \forall (i,j) \in \mathcal{E}^\mathrm{ext}, \\
    &\sum_{j \in \mathcal{V}^\mathrm{ext}}f_{ij} = 0, \quad &\forall i \in \mathcal{V}^\mathrm{ext}\setminus \{s,t\}, \\
    &\sum_{j \in \mathcal{V}^\mathrm{ext}}f_{sj} = d, \sum_{i \in \mathcal{V}^\mathrm{ext}}f_{it} = d
\end{align}
}
\end{prob}

\revb{The upper and lower bounds ensure that each order is allocated to a maximum of one driver, similar to constraint~\eqref{c1}. The weights $w_{ij}$ of the internal links are set to $1$, while the ones of all the others are set to $0$. The element $d$ can arbitrarily be either a variable or a parameter, representing the total number of drivers dispatched.} 

Fig.~\ref{fig:toy} depicts a toy example of the shareability network, \blue{and visually demonstrates how Problem~\ref{prob:one} is cast into Problem~\ref{prob:two}.}
In Fig.~\ref{fig:toy}(a), the green and red routes represent two subsequent orders. The first order (green) starts at 9 am and ends at 10 am, while the second order (red) starts at 11 am and ends at noon. The yellow route indicates that it takes 30 min to travel from the drop-off location of the first order to the pickup location of the second order. Since it is possible to travel from the destination of the first order to the origin of the second order before the pick-up time of the second order, we abstract the two orders with two squares connected with a link that goes from square 1 to square 2. The two trips correspond to two nodes in the vehicle-shareability network, and are connected by a ``connectivity link''. Fig. \ref{fig:toy}(b) shows a simplified shareability network without internal links. The simplified network allows to decrease the number of nodes and links in the network while maintaining  the connectivity information between order nodes.  Fig.~\ref{fig:toy}(c) shows the structure of the order nodes. The attributes $(1, 0, w)$ on the internal link are arranged in the following sequence: upper bound, lower bound, edge weight. 
Finally, Fig. \ref{fig:toy}(d) presents the extended shareability network with orders, drivers, virtual nodes, as well as internal links, connectivity links, and virtual links. Last, the required number of drivers is the overall flow from the source $s$ to the sink $t$. The solution to the maximum cost flow problem corresponds to the optimal dispatch. Various efficient solution techniques have been well investigated for minimum-cost flow problems \cite{cunningham1976network}, which guarantee computational tractability.

\footnotetext{Link attributes are in this order: upper bound, lower bound, edge weight.}
 
\section{Algorithms}
In the previous section we detailed how to formulate the VDP leveraging network flow models, i.e., formulating Problem~\ref{prob:one} into Problem~\ref{prob:two}. This section investigates different \revf{algorithmic} settings in which the maximum-cost flow optimization problem can be approached. \revb{First, in Section~\ref{sec:offline} and \ref{sec:online} we introduce two intuitive methodologies, offline and online, to solve the problem. We then present the formal definition of \textit{Sink Proximity} in Section~\ref{sec:sink} and show in Section~\ref{sec:sink-alg} how to leverage it to improve the previously proposed methodologies.}

\subsection{Offline Algorithm}\label{sec:offline}
\revb{In the offline approach, in which all the information is known \textit{a priori}, it is possible to compute a globally optimal solution, which can be used as a benchmark to assess the quality of online approaches. Here we solve the problem as a maximum cost network flow problem, using \emph{NetworkX} package in Python \cite{SciPyProceedings_11}. The construction of the network, and the result retrieving process are detailed in Algorithm~\ref{alg:offline}. }

\begin{algorithm}
\caption{Offline VDP}
\label{alg:offline}
\begin{algorithmic}[1]
\REQUIRE 
The \emph{FULL} set of orders $\mathcal{O}$ and  drivers $\mathcal{D}$ \\
\ENSURE 
Vehicle dispatching strategy $\mathcal{S}$.
\STATE Add all driver and order nodes to shareability network $\mathcal{G}$
\FOR{each driver/order pair $(d_i, o_j)$ in sets}
    \IF{$t_{i}^{d} + t_{ij} \leq t_{j}^{o}$}
    \STATE Add edge $(i,j)$
    \ENDIF
\ENDFOR
\FOR{each order/order pair $(o_i, o_j)$ in sets}
    \IF{$t_{i}^{o} + t_{ij} \leq t_{j}^{o}$}
    \STATE Add edge $(i,j)$
    \ENDIF
\ENDFOR
\STATE Solve the \revb{maximum} cost problem in the network $\mathcal{G}$
\STATE Recover driver order matching $\mathcal{S}$ from network flow
\STATE Return $\mathcal{S}$
\end{algorithmic}
\end{algorithm}

\subsection{Online Algorithm}\label{sec:online}
\revb{In online implementations, future orders and drivers are not fully known, especially ones far off in time. Thus, online} vehicle dispatch is more challenging than the offline counterpart due to limited information and forward-looking ability. In this section we leverage an RHC framework \cite{BorrelliBemporadEtAl2017} \rev{for online VDP}.

\begin{figure}[h]
\centering
\includegraphics[width=\linewidth]{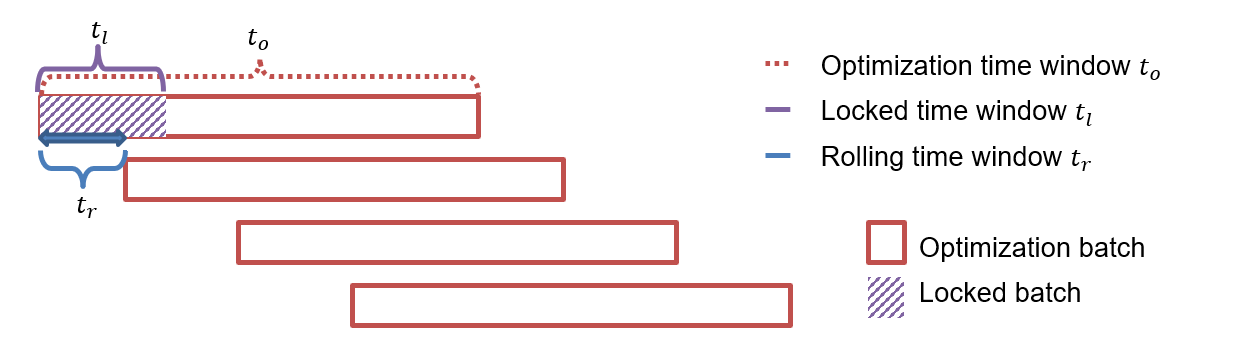}
\caption{Illustration of online receding horizon control algorithm \blue{structure}.}\label{fig:alg_draw}
\vspace{-6pt}
\end{figure}

Fig.~\ref{fig:alg_draw} \blue{illustrates the RHC framework of} the online algorithm\blue{\footnote{Due to page limitations, algorithmic implementations are in the appendix.}}. \revb{In RHC, the VDP problem is solved iteratively in a short optimization time window, $t_o$\footnote{The order information in $t_o$ is known by scheduling or by forecasting.}. In each step, a vehicle is dispatched for a locked time window $t_l$, after which the time horizon rolls by $t_r$}\footnote{In the setting of this paper, $t_o \ge t_l \ge t_r$ always holds. \revb{This is to avoid the ``dead zone'' at the beginning of each optimization time window, where experience delays as vehicles are dispatched to each request's start location}}.

\blue{D}ue to the \blue{limited forward-looking horizon}, we ignore the future order connectivity in the shareability network. This leads to a greedy suboptimal solution for the online algorithm \cite{mpc_1999}.

\subsection{Sink Proximity}\label{sec:sink}
To address the \revb{suboptimality} in \blue{the online RHC algorithm}, we propose a new \revb{network science methodology that} overcomes the shortsightedness \revb{of the RHC framework, \emph{sink proximity}.}

The goal of this methodology is to create an attribute for each node (order) of the network that \revb{captures future connectivity of the network from this node while being} easy and efficient to compute. \revb{In other words, this attribute captures the importance of each order from a network-wide perspective and can be leveraged in the online VDP to compensate for the lack of future information.} 
\revb{We introduce the formal definition of \emph{sink proximity}.}
\begin{definition}
Given a shareability network $\mathcal{G} = (\mathcal{V}, \mathcal{E})$ with $\mathcal{V} := \mathcal{O} \cup t$, and the set of all the paths from node $v_i$ to $t$, $ P(v_i,t)$, Sink Proximity of node $v_i$ is defined as 
\begin{equation}
    \mathrm{SP}_{v_i} = \max\quad{P}(v_i,t).
\end{equation}
\end{definition}

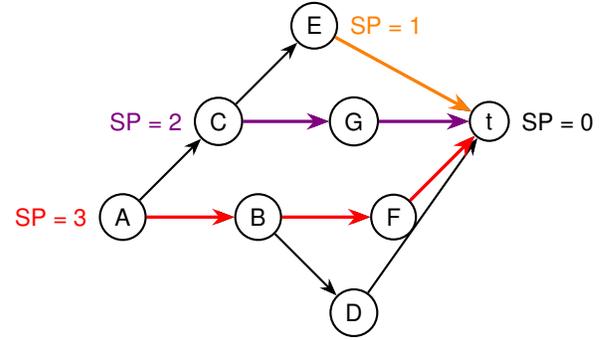
\begin{figure}[bt]
    \centering
\begin{tikzpicture}[
    ->,
    >=Stealth,
    node distance=2cm and 1.5cm,
    thick,
    every node/.style={draw, circle},
    every edge/.style={draw, thick},
    scale=0.9,
    transform shape
]

\node (A) [label=left:\textcolor{red}{SP = 3}] {A};
\node (B) [right of=A] {B};
\node (C) [above right of=A, label=left:\textcolor{violet}{SP = 2}] {C};
\node (D) [below right of=B] {D};
\node (F) [right of=B] {F};
\node (G) [right of=C] {G};
\node (E) [above right of=C, label=right:\textcolor{orange}{SP = 1}] {E};
\node (T) [right of=G, label=right:{SP = 0}] {t};

\draw (A) -- (B);
\draw (A) -- (C);
\draw (B) -- (D);
\draw (B) -- (F);
\draw (C) -- (E);
\draw (C) -- (G);
\draw (D) -- (T);
\draw (E) -- (T);
\draw (F) -- (T);
\draw (G) -- (T);

\draw[red, very thick] (A) -- (B);
\draw[red, very thick] (B) -- (F);
\draw[red, very thick] (F) -- (T);
\draw[violet, very thick] (C) -- (G);
\draw[violet, very thick] (G) -- (T);
\draw[orange, very thick] (E) -- (T);

\end{tikzpicture}
\vspace{-0.2cm}
\caption{Schematic representation of a simplified shareability network with the SP attribute \blue{of Node A, C, E, $t$. The color denotes the longest route and its corresponding SP value.} }
\label{fig:tikzex}
\vspace{-10pt}
\end{figure}

\revb{In a shareability network, the sink proximity} of an \rev{order} is the maximum number of {orders that} a driver can be matched with {after serving} this order. Fig.~\ref{fig:tikzex} provides a schematic representation of this concept, \blue{with each color denoting the longest route and its corresponding SP value from node A (red), C (purple), E (orange), and the terminal node $t$ (SP = 0).}

In addition, \emph{sink proximity} can be efficiently computed using Proposition \ref{prop:1}. 

\begin{prop} \label{prop:1}
    Finding the maximum number of orders a driver can take in a shareability network can be expressed as a single-source DAG longest path problem. 
\end{prop}

\begin{proof}
Given a shareability network $\mathcal{G}$, each edge represents the connectivity between a driver and an order or between orders. A driver can start from any node and take orders along any edge, as long as it has not been used before. The driver’s objective is to maximize the total number of orders taken in the future, or order-to-go. This problem is equivalent to a single source DAG longest path problem. 
Note that any feasible solution for the driver corresponds to a path in $\mathcal{G}$, and vice versa. Moreover, the total number of orders taken by the driver is equal to the length of the corresponding path. Therefore, maximizing the number of orders is equivalent to maximizing the length of the path. As $\mathcal{G}$ is a shareability network, the acyclic nature of time guarantees that it has no cycles \cite{vazifeh2018addressing}. 
This means that it is possible to solve the problem in linear time by processing the vertices in topological order, usually, with a complexity of \rev{$\mathcal{O}(|\mathcal{V}|+|\mathcal{E}|)$}. 
\end{proof}

Computing sink proximity of a node requires full knowledge of the network. However, it is possible to efficiently estimate this attribute using a data-driven approach \revb{as shown in Appendix~\ref{sec:sp-forcast}}. 

\subsection{\revb{Sink Proximity-Aided Algorithm}}\label{sec:sink-alg}

In this section, we demonstrate how to leverage \emph{sink proximity} to improve the quality of the solution obtained by Algorithm~\ref{alg:onlineVDP}. The attribute \emph{sink proximity} of a node indicates the number of additional nodes that can be visited after that particular node. It is reasonable to assume that the nodes with higher \emph{sink proximity} should have higher priority. Thus, in the context of this work, given the \emph{sink proximity} value of an order, we assign its value to the weight of the internal link of that order. 
The detailed algorithm that leverages sink proximity in an RHC framework is presented in Algorithm~\ref{alg:online-SP}, called RHC-SP. 
Compared to Algorithm~\ref{alg:onlineVDP}, the only difference lies in the additional ``for loop'' in lines 4 to 7, in which the sink proximity attribute of each node \revb{is used to modify the weight of the internal links for order nodes (illustrated in Fig. \ref{fig:toy} (c))}. 
Note that in the extended shareability network used as input in Algorithm~\ref{alg:online-SP}, \revb{still} only the internal links can have non-zero weights because the successful fulfillment of an order is only represented by one unit of flow in the internal link.

\begin{algorithm}[t]
\caption{Online VDP with RHC-SP}\label{alg:online-SP}
\begin{algorithmic}[1]
\REQUIRE 
A set of orders $\mathcal{O}$ and a set of drivers $\mathcal{D}$ in an optimization time window $t_o$. 
A rolling time window $t_r$, and a locked time window $t_l$ where decisions are locked. 
\ENSURE 
Vehicle dispatching strategy $\mathcal{S}$.
\STATE Initialize $t \gets t_\text{start}$
\WHILE{the system is running}
    \STATE Get order set $\mathcal{O}_t$ and driver set $\mathcal{D}_t$ in $[t, t + t_o]$
    \FOR{each order node $o$ set $\mathcal{O}$}
        \STATE Estimate sink proximity $\rev{\mathrm{SP}_o}$
        \STATE Assigned $\rev{\mathrm{SP}_o}$ as the weight to \rev{interal} edge of $o$
    \ENDFOR
    \STATE Construct shareability network $\mathcal{G}_t$
    \STATE Solve the \revb{maximum} cost problem in the network $\mathcal{G}_t$
    \STATE Recover driver order matching $\mathcal{S}_t$ from network flow
    \STATE Apply actions for matching in $[t, t + t_l]$ and add to $\mathcal{S}$
    \STATE Update order set $\mathcal{O}_t$
    \STATE Update driver set $\mathcal{D}_t$
    \STATE Update $t = t + t_r$
\ENDWHILE
\STATE Return $\mathcal{S}$
\end{algorithmic}
\end{algorithm}

\section{Numerical Study}
In this section we conduct a case study of Manhattan, NYC, USA. 
For the orders we use the NYC Taxi and Limousine Commission (TLC) trip record data \cite{NYC_taxi}. The dataset contains the start time, start location (as an area code), end time, and end location (as an area code) of each taxi order. The city is divided into 260 non-overlapping areas \revb{during data processing. Thereafter, the peak hours of two mornings (8:00 to 10:00 am) on June 1 and 2, 2022, are selected. There are} 11,483 and 11,825 requests in these two days, respectively. \blue{The code is available at: \url{https://github.com/TIVV424/SPMoD}}.

Table \ref{tab:para} summarizes the parameters used in the numerical experiment. 
The algorithm's performance is evaluated based on the RSR and running time. \revb{}

\begin{table}[h]
    \centering
    \caption{Parameter settings in numerical experiment}\label{tab:para}
    \begin{tabular}{ccc}
    \toprule
    Parameter &  Note & Value\\ 
    \hline
    $t_o$     & Optimization time window & 10, 20, 30, 60 min\\
    $t_r$     & Rolling time window & 5, 10, 20, 30 min\\
    $t_l$     & Locked time window & 8, 10, 15, 20, 30 min\\
    $n_D$     & Number of drivers & 2000\protect\footnotemark \\
    $t_\text{void}$     & Driver idle time & 30 min\\
    \bottomrule
    \end{tabular}
\end{table}
\vspace{-12pt}
\footnotetext{The driver-to-order ratio is \rev{approximately} 0.2}

\subsection{\revb{Analyzing Algorithm Performance}}

\revb{This section showcases the performance of our proposed approach. First, it is crucial to forecast the values of \emph{sink proximity} (see Appendix \ref{sec:sp-forcast}) for each node, and modify the weights of the internal links of the orders, as described in Section~\ref{sec:sink-alg}. For a given simulation, the optimal objective of the maximum cost network flow problem, from which it is possible to retrieve the optimal objective of the VDP problem, i.e., the RSR, is given by the offline approach, Algorithm \ref{alg:offline}, which yields global optimality.}

\begin{figure*}[t]
\vspace{-6pt}
    \centering
    \includegraphics[trim={5 10
0 12}, width= \linewidth]{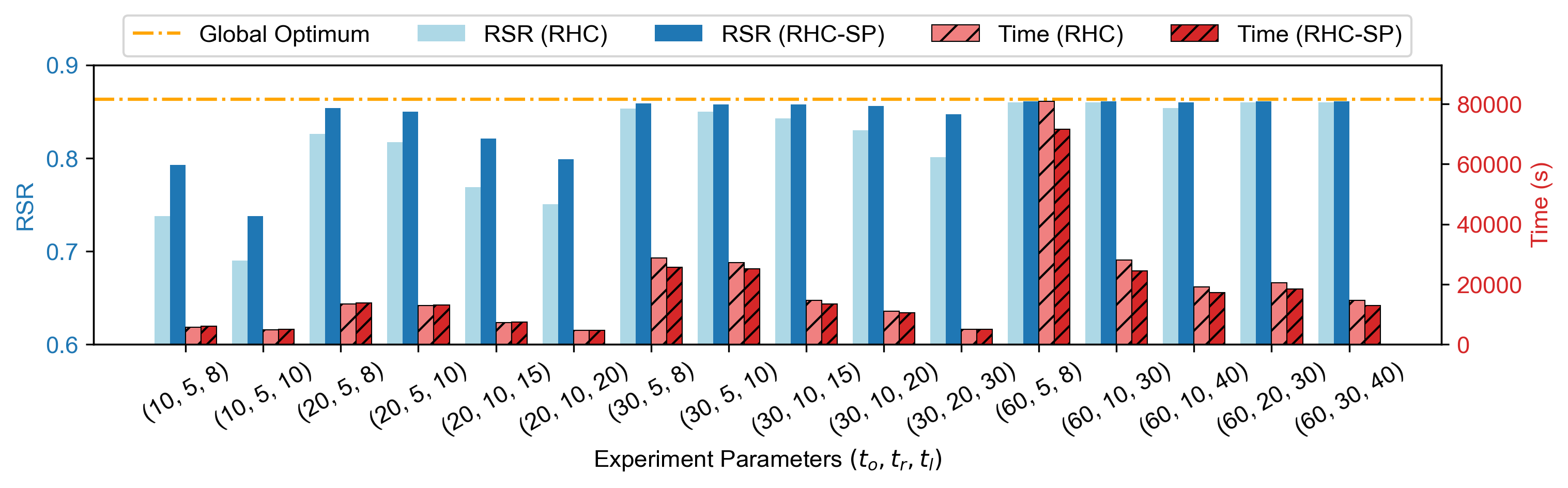}
    \caption{Comparison of RSR \blue{(left axis), i.e., objective of Problem~\ref{prob:one}, } and computation time \blue{(right axis)} for online algorithms with different parameter settings. \blue{The global optimal solution is computed by solving the same problem with the offline algorithm.}}\label{fig:time}
\vspace{-8pt}
\end{figure*}

Fig.~\ref{fig:time} \blue{compares} two algorithms, RHC with and without \emph{sink proximity}. We also conduct a sensitivity analysis of several parameters, namely the optimization time window $t_o$, the locked time window $t_l$, and the rolling time window $t_r$. 
\blue{Fig.~\ref{fig:time} demonstrates that RHC-SP algorithm consistently outperforms standard RHC in terms of RSR.} It is important to note that the complexity for both algorithms with the same parameter settings is the same, and it takes approximately comparable time to compute. This is because the original structure of Problem~\ref{sec:network_formulation} remains unchanged, regardless of the usage of \emph{sink proximity}. Thus, \blue{RHC-SP permits the selection of more aggressive parameters}, such as a shorter optimization interval. This is thanks to a more efficient computation, which is vital for ride-hailing platforms that have to operate online.

The results indicate that the proposed RHC-SP algorithm reduces the run time by 4.55\%\footnote{The reported number demonstrates that the algorithm does not compromise computational efficiency, but it does not indicate an increase in computational efficiency.}, and increases the \revf{RSR} by 1.95\% on average across all parameter settings tested. 

\begin{table}[h]
\vspace{-6pt}
\centering
\caption{\reva{Objective value improvement of RHC-SP compared to RHC, with different parameter settings.}}\label{tab:rsr_online}
\begin{tabular}{cccc}
\toprule
$t_o$ & $t_r$ & $t_l$ & RSR improvement\\
\hline
10 & 5 & 8 & \cellcolor{green!25}7.41 \% \\
10 & 5 & 10 & \cellcolor{green!20}6.98 \% \\
20 & 5 & 8 & \cellcolor{yellow!25}3.35 \% \\
20 & 5 & 10 & \cellcolor{yellow!25}4.00 \% \\
20 & 10 & 15 & \cellcolor{green!20}6.80 \% \\
20 & 10 & 20 & \cellcolor{green!20}6.47 \% \\
30 & 5 & 8 & \cellcolor{red!25}0.62 \% \\
30 & 5 & 10 & \cellcolor{red!25}0.94 \% \\
30 & 10 & 15 & \cellcolor{yellow!15}1.76 \% \\
30 & 10 & 20 & \cellcolor{yellow!25}3.13 \% \\
30 & 20 & 30 & \cellcolor{yellow!25}5.75 \% \\
60 & 5 & 8 & \cellcolor{red!25}0.15 \% \\
60 & 10 & 30 & \cellcolor{red!25}0.17 \% \\
60 & 10 & 40 & \cellcolor{red!25}0.67 \% \\
60 & 20 & 30 & \cellcolor{red!25}0.12 \% \\
60 & 30 & 40 & \cellcolor{red!25}0.18 \% \\
\bottomrule
\end{tabular}
\vspace{-6pt}
\end{table}

Table \ref{tab:rsr_online} illustrates the performance comparison between the RHC-SP and RHC Algorithms \revb{in terms of RSR improvement}. The RHC-SP algorithm consistently outperforms the RHC algorithm, especially when $t_o$ is small. To enhance clarity, the magnitude of improvement is visually represented using color coding: \blue{green indicates large gains, while red indicates small ones.}

\begin{figure*}[tbh]
    \centering     
    {\includegraphics[width=\linewidth]{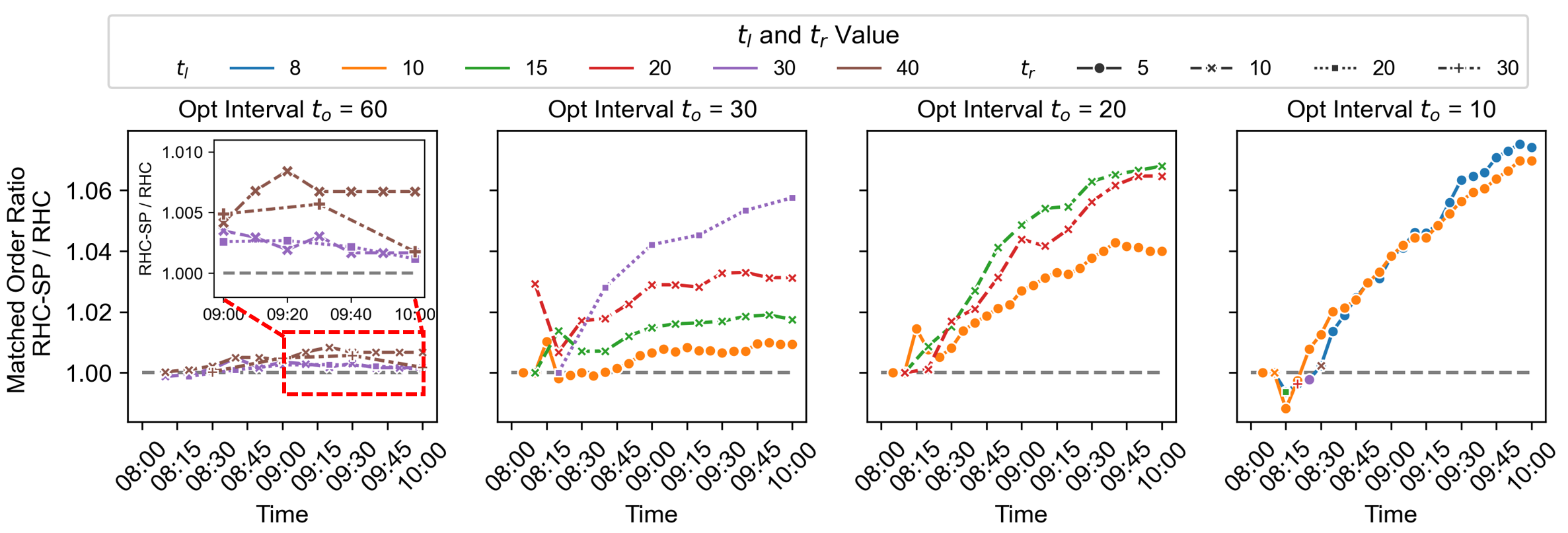}}
    \caption{\blue{Comparison of rolling request service rate (Problem~\ref{prob:one} objectivev, RSR) using RHC-SP and RHC across parameter settings. Ratios above 1 indicate that the proposed RHC-SP algorithm outperforms RHC.}}\label{fig:match}
\vspace{-12pt}
\end{figure*}

To gain deeper insights into the results, We examine the rolling RSR through the time horizon, as shown in Fig.~\ref{fig:match}, which depicts the \revf{rolling} ratio of the \revf{RSR, up to that point in time, i.e.,} at the end of each rolling iteration, between the two algorithms. When the lines are above the gray dashed line (ratio of 1), it indicates that RHC-SP outperforms the RHC algorithm up to that point in time. 
{Among all parameter settings, RHC-SP with the $t_o = 20, t_r = 10, t_l = 20$ parameter combination provides the most performance improvement w.r.t. the standard RHC framework. }

We highlight that, RHC-SP, at the end of the simulation, always outperforms the standard RHC in terms of \gls{RSR}. The benefit of  RHC-SP is generally more significant when the optimization time window is smaller and the rolling/locked windows are larger, i.e., when future information is limited. This indicates that the algorithm is especially valuable when computational power is restricted and a larger rolling time window and locked time window are necessary for quick dispatch and decision-making. 
Interestingly, we highlight that in Fig.~\ref{fig:match}, the cumulative RSR with parameter $t_o=10$ does not constantly outperform RHC during the whole simulation time, even if at the end it does. This result confirms that the sink proximity attribute allows the RHC algorithm to not be myopic because of the horizon limit, allowing to overall outperform, even if, in the short-term, it might underperform because of the additional future information. From a physical point of view, the result can be interpreted as a preventive realignment to meet future orders by sacrificing current ones. To summarize, sink proximity complements the missing information on the network's future structure and evolution \rev{when} solving an online problem with RHC.  

\subsection{Discussions}
\revb{RL has recently been applied to various combinatorial problems,} including the VDP. Although this work is orthogonal to RL, it is worth exploring \revb{how the proposed approach could be integrated into an RL framework.} Notably, the authors in~\cite{SchmidtGammelliEtAl2024,SinghalGammelliEtAl2024} show that, in a similar setting compared to \revb{ours}, RL outperforms in terms of computation time with respect to an RHC framework but it under performs in terms of overall profit generated by the fleet operator. The RHC framework, in turn, is sub-optimal w.r.t. the ``oracle'' case, i.e., the equivalent of our offline approach. 

In online VDP, the most significant challenge arises from uncertainty regarding future demand. Rather than attempting to predict the precise timing and location of future orders as in many CO works, the concept of sink proximity allows us to estimate aggregated network information. The latter is a simpler task that enhances the reliability of the predicted data. The concept of sink proximity also resembles value functions in RL. While the value function captures the expected future reward of a given state, sink proximity captures the projected value of a given order node. In future work, it would be interesting to investigate if our proposed approach can be integrated in an RL framework, and to what extent it would be beneficial to the quality of the solution obtained.

\section{{Conclusions}}

\blue{This paper examines how to dispatch vehicles more effectively in ride-hailing platforms to ease traffic flow, elevate service quality, and strengthen the resilience of urban mobility systems.}
\blue{We introduce} a novel approach to improve the request service rate on ride-hailing platforms by incorporating network science \blue{measures} into the shareability network. Sink proximity, a \blue{measure for} the longest path distance to the sink node within a network flow context, differs from existing \blue{measures} by focusing on the global behavior of the network rather than local features of individual nodes or regions. By converting the problem into its equivalent form of finding the longest path in a DAG, we show that sink proximity can be efficiently calculated and predicted from order node information in a shareability network.

Building upon this, we propose a new online dispatch algorithm, RHC-SP, which assigns edge weights to the nodes based on their sink proximity. This allows the algorithm to incorporate network expansion information, even in a limited-time, partial network. By integrating this into the online optimization process, the proposed algorithm improves performance, increasing the number of matched orders by up to 7\% compared to a standard RHC framework, without additional computational overhead. The inclusion of sink proximity as a weight in the dispatch problem resembles the terminal constraints used in RHC problems, addressing the suboptimality that can arise when the optimization horizon is limited.

Future work could explore scenarios with more volatile request distributions or investigate a sliding sink node approach to simulate infinite-horizon problems. Additionally, analyzing sink proximity in other dynamic network applications holds promising potential for further research. \blue{While a direct numerical comparison with other VDP algorithms is not possible due to differences in problem assumptions and modeling frameworks, another promising direction for future work is to incorporate SP into other online algorithms and evaluate its impact on performance.}
\vspace{-10pt}
\appendices



\section{Online VDP}

\blue{We present Algorithm~\ref{alg:onlineVDP} as the online version of Algorithm~\ref{alg:offline} using RHC techniques.
In this algorithm, we detail how to iteratively construct the network and simplify the representation of the network construction process. Steps 1 to 11 in Algorithm~\ref{alg:offline} are reduced to Step 4 in Algorithm~\ref{alg:onlineVDP}.}

\begin{algorithm}
\caption{Online VDP with RHC}\label{alg:onlineVDP}
\begin{algorithmic}[1]
\REQUIRE 
A set of orders $\mathcal{O}$ and a set of drivers $\mathcal{D}$ in an optimization time window $t_o$. 
A rolling time window $t_r$, and a locked time window $t_l$ where decisions are locked. 
\ENSURE 
Vehicle dispatching strategy $\mathcal{S}$.
\STATE Initialize $t \gets t_\text{start}$
\WHILE{the system is running}
    \STATE Get order set $\mathcal{O}_t$ and driver set $\mathcal{D}_t$ in $[t, t + t_o]$
    \STATE Construct shareability network $\mathcal{G}_t$
    \STATE Solve the maximum cost problem in the network $\mathcal{G}_t$
    \STATE Recover driver order matching $\mathcal{S}_t$ from network flow
    \STATE Apply actions for matching in $[t, t + t_l]$ and add to $\mathcal{S}$
    \STATE Update order set $\mathcal{O}_t$
    \STATE Update driver set $\mathcal{D}_t$
    \STATE Update $t = t + t_r$
\ENDWHILE
\STATE Return $\mathcal{S}$
\end{algorithmic}
\end{algorithm}
\vspace{-10pt}

\section{\revb{Sink Proximity Forecasting}}\label{sec:sp-forcast}

\revb{In online dispatch with receding horizon, the network structure remains unknown after the optimization window. Thus, forecasting is necessary to estimate the sink proximity value over a period longer than the time window. We leverage data-driven prediction methods to estimate the value of sink proximity, where the inputs include order start area, start time, end area, and end time or the orders.}

\subsection{Time-standardized Sink Proximity}\label{sec:tssp}

The sink proximity metric measures the step-wise reachability of nodes to the sink in a network with a longer horizon and captures information about how the shareability network expands. To illustrate this concept, in the context of the shareability network, an order node with a higher degree of sink proximity is more probable to be an order in a demand-heavy area during peak hours. 

If the end time of a node is closer to the end time of the network construction horizon, the sink proximity will naturally be smaller. Thus, {to mitigate this phenomenon,} the time-standardized sink proximity (TSSP) is defined as follows:

\begin{definition}
For each node $v_i$ in a shareability network $\mathcal{G}= (\mathcal{V},\mathcal{E})$, the time-standardized sink proximity is computed as the sink proximity divided by the time it takes to reach the sink node, with a buffer. Thus it is defined as 
\begin{equation}
    \mathrm{TSSP}_\rev{v_i} = \frac{\mathrm{SP}_\rev{v_i}}{t_e + t_\text{buffer}}.
\end{equation}
\end{definition}

These measurements quantitatively represent the importance of each order node in the network. Note that if a sink node has a time attribute $t$, and the time it takes to reach the sink node is a small time period  $ t_e$, most of the orders with an end time $t_{i}^{d} \geq t-t_e$ would have a very small sink proximity value. Thus, computing TSSP allows for eliminating the impact of the sink node time attribute, facilitating the prediction of the network metric.

\subsection{\revb{Sink Proximity Forecasting Model}}

We use $\epsilon$-SVR~\cite{smola2004} with radial basis function kernel to predict the value of TSSP, and tune the hyperparameters\footnote{See ``https://scikit-learn.org/1.5/modules/generated/sklearn.svm.SVR.html'' for parameter explanations.}, shown in Table~\ref{tab:svr}, using grid search and cross-validation. Then, we reverse the time standardization and round the value to recover the sink proximity for each order, which \revb{are integers} by definition. 

\begin{table}[h]
\vspace{-10pt}
    \centering
    \caption{Grid search range and best parameters}\label{tab:svr}
    \begin{tabular}{ccc}
    \toprule
    Parameter &  Grid search ranges & Best value\\ 
    \hline
    kernel     & [poly, rbf, sigmoid] & rbf\\
    L2 regularization parameter   & [100, 200,300, 400, 500] & 300\\
    gamma    & [0.001, 0.01, 0.1] & 0.01\\
    epsilon  & [0.01, 0.05, 0.1, 0.2, 0.5] & 0.1\\
    \bottomrule
    \end{tabular}
\vspace{-10pt}
\end{table}
\vspace{-10pt}

\subsection{\revb{Sink Proximity Forecasting Results}}

\begin{figure*}[t]
    \centering
    \includegraphics[width= \linewidth,trim={0 1cm 0 0}]{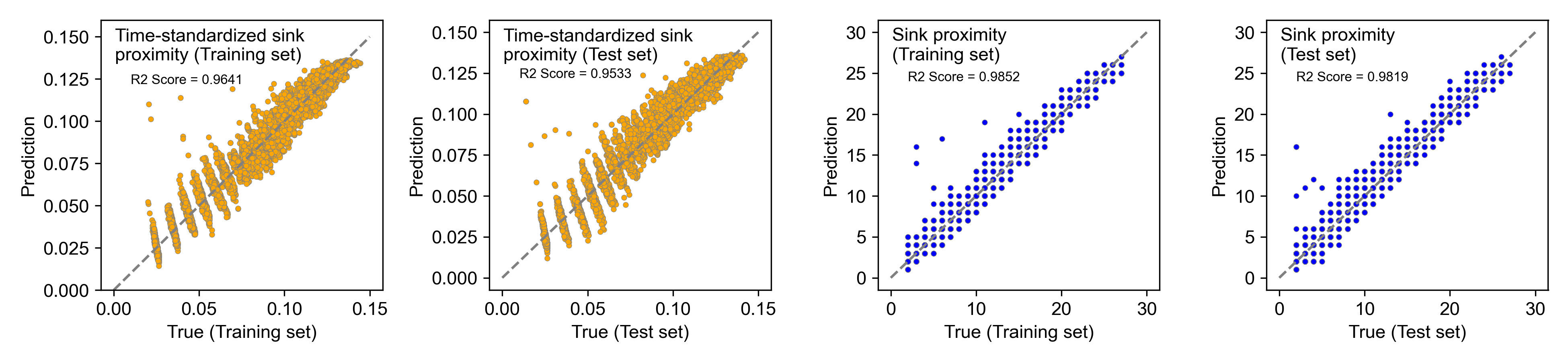}
    \vspace{-6pt}
    \caption{Prediction accuracy of \revb{TSSP and SP} \reva{in both training and testing sets} for the case study of Manhattan, NYC. Data used for training is the morning peak hour of June 1st, 2022. Data used for validation is the morning peak hour of June 2nd, 2022. } \label{fig:pred}
\vspace{-10pt}
\end{figure*}

To train the model to forecast sink proximity, we leverage data of the morning peak hour dataset of June 1st, 2022, and test on the same period on the following day. \revb{The buffer time for TSSP calculation is selected to be 20 minutes.} Fig.~\ref{fig:pred} illustrates the prediction accuracy of sink proximity \reva{in training and testing dataset}. We highlight that for an order node that is closer to the sink in time with a low sink proximity, the prediction accuracy is also lower. This is due to the relative magnitude of the value, as shown in the lower-left corner of Fig.~\ref{fig:pred}, where the strip pattern is displayed. However, this lower accuracy can be improved by reversing the standardization to recover the integral sink proximity. \revb{After rounding, the prediction accuracy increased. 
Using the SVR model, the predicted TSSP achieves an $R^2$ score of 0.95, while the recovered SP after rounding reaches an $R^2$ score of 0.98.}
Using the SVR model, the predicted sink proximity achieves an $R^2$ score of 0.985 in the training set, and 0.982 in the testing set. 

\vspace{-10pt}

\bibliographystyle{IEEEtran}  %
\bibliography{refs}

\vspace{-33pt}
\begin{IEEEbiography}[{\includegraphics[width=1in,height=1.25in,clip,keepaspectratio]{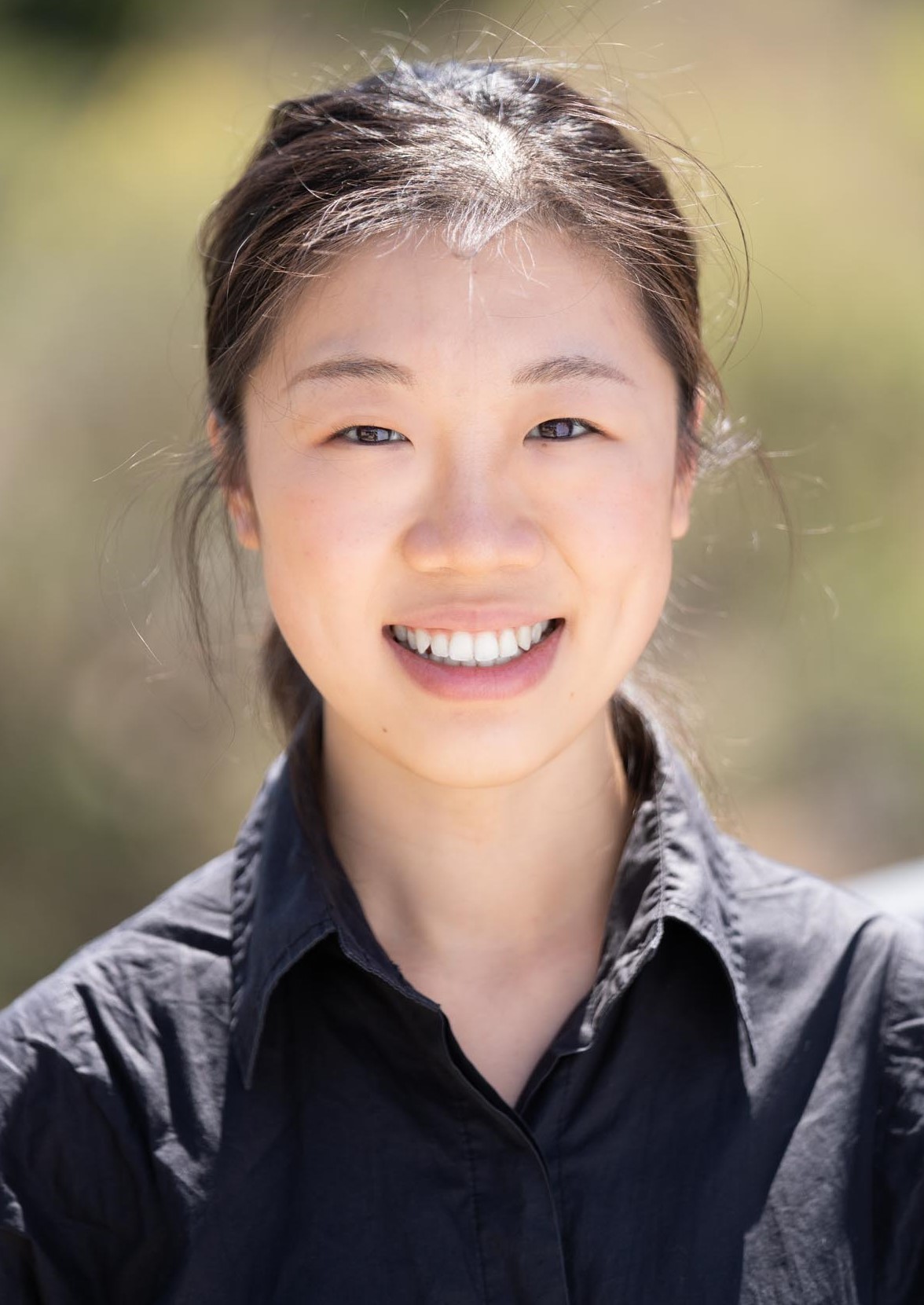}}]{Ruiting Wang} is a Postdoctoral Fellow at KTH Royal Institute of Technology. She received her Ph.D. degree in Systems Engineering from the University of California, Berkeley in 2025, and got her B.S. degree in Building Energy Engineering from Tsinghua University in 2020. Her research combines ideas from optimization, control, economics, and learning to develop algorithms for socio-technical decision-making problems in future energy-mobility systems, from urban energy infrastructure, to intelligent transportation networks, to EV markets. 
\end{IEEEbiography}

\vspace{-10pt}

\begin{IEEEbiography}[{\includegraphics[width=1in,height=1.25in,clip,keepaspectratio]{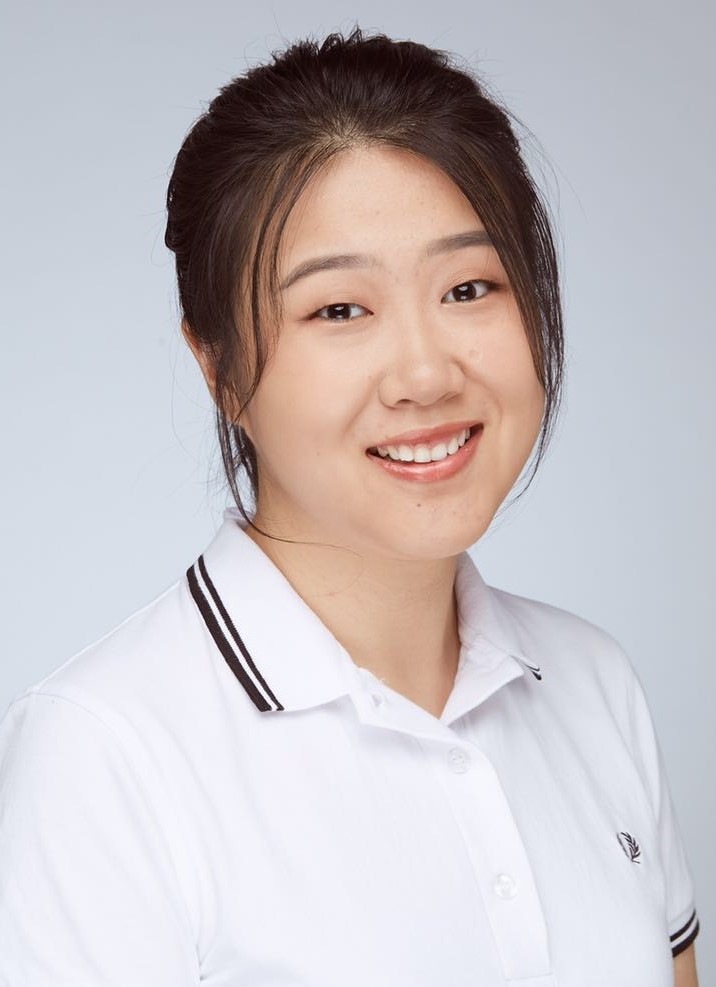}}]{Jiaman Wu} is an Assistant Professor in Data Science at City University of Hong Kong. She got her Ph.D. degree in Systems Engineering in the Department of Civil and Environmental Engineering, UC Berkeley. Before that, she received her bachelor's degree in geographical monitoring and census from the School of Remote Sensing and Information Engineering, Wuhan University, and her master's degree in computer science and technology from the Institute for Interdisciplinary Information Sciences, Tsinghua University. Her research interests include operation and planning for the power system and transportation system.
\end{IEEEbiography}

\vspace{-10pt}

\begin{IEEEbiography}[{\includegraphics[width=1in,height=1.25in,clip,keepaspectratio]{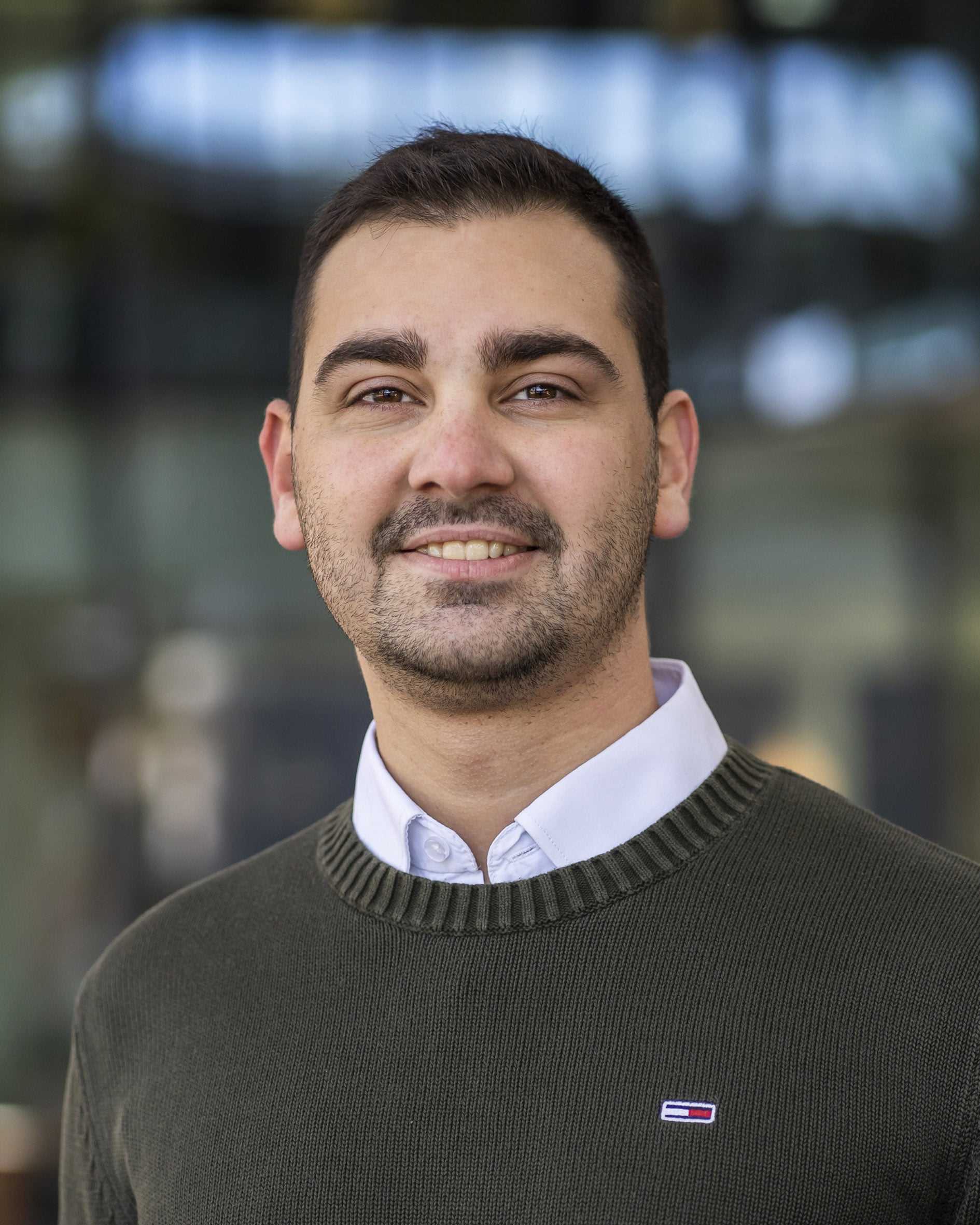}}]{Fabio Paparella} obtained his Ph.D. from the Control Systems Technology (CST) section at Eindhoven University of Technology, the Netherlands. He studied mechanical engineering at Politecnico di Milano, Italy, where he received his Bachelor's degree in 2017 and his Master's cum laude in 2020 with a thesis in collaboration with NASA Jet Propulsion Laboratory, California, USA. His research interests include mobility-on-demand, smart mobility, and optimization. In 2024 he was a visiting scholar at University of California, Berkeley.
\end{IEEEbiography}

\vspace{-10pt}

\begin{IEEEbiography}[{\includegraphics[width=1in,height=1.25in, clip,keepaspectratio]{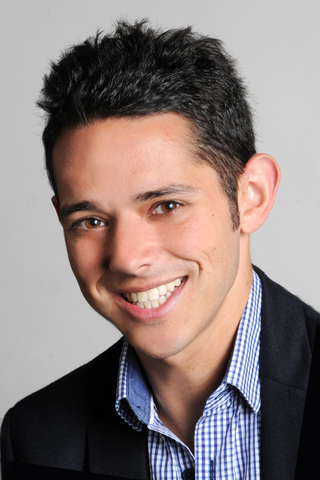}}]{Scott J. Moura} is a Full Professor in Civil \& Environmental Engineering and Director of the Energy, Controls, \& Applications Lab (eCAL) at the University of California, Berkeley. 
He received a B.S. degree from the University of California, Berkeley, CA, USA, and the M.S. and Ph.D. degrees from the University of Michigan, Ann Arbor, in 2006, 2008, and 2011, respectively, all in mechanical engineering. From 2011 to 2013, he was a Post-Doctoral Fellow at the Cymer Center for Control Systems and Dynamics, University of California, San Diego. In 2013, he was a Visiting Researcher at the Centre Automatique et Systèmes, MINES ParisTech, Paris, France. His research interests include control, optimization, and machine learning for batteries, electrified vehicles, and distributed energy resources.

Dr. Moura is a recipient of the National Science Foundation (NSF) CAREER Award, Carol D. Soc Distinguished Graduate Student Mentor Award, the Hellman Fellowship, the O. Hugo Shuck Best Paper Award, the ACC Best Student Paper Award (as advisor), the ACC and ASME Dynamic Systems and Control Conference Best Student Paper Finalist (as student and advisor), the UC Presidential Postdoctoral Fellowship, the NSF Graduate Research Fellowship, the University of Michigan Distinguished ProQuest Dissertation Honorable Mention, the University of Michigan Rackham Merit Fellowship, and the College of Engineering Distinguished Leadership Award.
\end{IEEEbiography}

\vspace{-10pt}

\begin{IEEEbiography}
[{\includegraphics[width=1in,height=1.25in,clip,keepaspectratio]{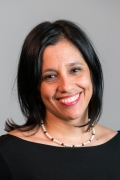}}]{Marta C. Gonzalez} is a Full Professor of City and Regional Planning, Civil and Environmental Engineering at the University of California, Berkeley, and a Physics Research faculty in the Energy Technology Area (ETA) at the Lawrence Berkeley National Laboratory (Berkeley Lab). Prior to joining Berkeley, Dr. Gonzalez worked as an Associate Professor of Civil and Environmental Engineering at MIT, a member of the Operations Research Center and the Center for Advanced Urbanism. She is a member of the scientific council of technology companies such as Gran Data, PTV and the Pecan Street Project consortium.

Dr. Gonzalez received the Licentiate degree in physics from the Universidad Simon Bolivar, in 1999, the Magister Sc. degree in physics from Central University of Venezuela, in 2001, and the PhD (Dr. rer. nat) degree in physics from Stuttgart University at in 2006. She has more than 60 publications, and her work has been published in the Nature, the Science, the Nature Physics, the Physics A, the Journal of the Royal Society Interface, the Physical Review Letters, the Scientific Reports, the Transportation Research Part C: Emerging Technologies, the Data Mining and Knowledge Discovery, among many others.
\end{IEEEbiography}


\vfill

\end{document}